\documentclass[aps,pre,reprint,floatfix,10pt]{revtex4-1}
\usepackage[dvipsnames]{xcolor}
\usepackage{graphicx,mathtools,amsmath,amssymb,tabularx,hyperref,bm}
\begin{document}
	\newcommand{\arXiv}[1]{{\tt \href{http://arXiv.org/abs/#1}{arXiv:#1}}}

\newcommand{\ee}[1]{\mathrm{e}^{#1}}
\newcommand{\ii}{\mathrm{i}}
\newcommand{\dd}{\mathrm{d}}
\newcommand{\tr}{\mathrm{tr}}
\newcommand{\id}{\mathbf{1}}
\newcommand{\APD}{\mathrm{APD}}
\newcommand{\DI}{\mathrm{DI}}
\newcommand{\EQ}[1]{\mathrm{EQ}_{#1}}
\newcommand{\diag}[1]{\mathrm{diag}(#1)}
\DeclarePairedDelimiter{\ceil}{\lceil}{\rceil}
\DeclarePairedDelimiter{\floor}{\lfloor}{\rfloor}
\renewcommand{\Re}{\mathrm{Re}}
\renewcommand{\Im}{\mathrm{Im}}
\newcommand{\erf}{\mathrm{erf}}
\newcommand{\wid}{x_s}
\newcommand{\pos}{\theta}
\newcommand{\amp}{A}

\renewcommand{\vec}[1]{\mathbf{#1}}
\newcommand{\op}[1]{\mathcal{#1}}

\newcommand{\CM}[1]{{#1}}

\newcommand{\vX}[1]{\check{\vec{X}}(x-\theta; x_s)}
\newcommand{\vmu}[1]{\check{\mu}_{#1}(s; \theta, x_s)}
\newcommand{\Us}{\check{U}(s; \theta, x_s)}
\newcommand{\vPhi}[1]{\check{\vec{\Phi}}_{#1}(x-s; s)}

\title{\CM{Predicting effective q}uenching of stable pulses in slow-fast excitable media.}

\author{Christopher D. Marcotte}
\affiliation{Department of Computer Science, Durham University, Durham, UK, DH1 3LE}~\thanks{Corresponding author: christopher.marcotte@durham.ac.uk}

\date{\today}

\begin{abstract}
We develop \CM{a} linear theory for the prediction of excitation wave quenching---the construction of minimal perturbations which return stable excitations to quiescence---for localized pulse solutions \CM{in} models of excitable media.
The theory \CM{accounts} for an additional \CM{equivariance compared to the homogeneous ignition problem}, and \CM{thus requires} a reconsideration of heuristics for choosing optimal reference states from their group representation.
We compare predictions made with the linear theory to direct numerical simulations across a family of perturbations and assess \CM{their} accuracy \CM{for several models} with distinct stable excitation structures.
We find that the theory achieves qualitative predictive power with only the effort of \CM{continuing} a \CM{scalar} root, and achieves quantitative predictive power in many circumstances.
Finally, we compare the computational cost of our prediction technique to other numerical methods for the determination of transitions in extended excitable systems.
\end{abstract}

\keywords{excitable systems, quenching, eigen-analysis, reaction-diffusion models, cardiac dynamics}

\maketitle

\section{Introduction}

\CM{In the simplest case, ventricular and atrial cardiac tissues are modelled as a continuous excitable medium, and fibrillation is understood as a chaotic excitation driven by the presence of several interacting electrical excitation waves.
Since fibrillation lowers the efficacy of the heart at moving oxygenated blood, it is imperative that the normal rhythm is restored, i.e. through defibrillation.
During defibrillation, an electric field is imposed across the tissue to recruit new excitations from repolarized -- excitable -- tissue regions.
New excitations destructively interfere with some proportion of the extant excitation wave fronts, thus reducing the proportion of tissue which is fibrillating.
To completely halt fibrillation, multiple pulses are applied with a particular delay, which lowers the energy requirement both per-pulse and overall, while maintaining efficacy~\cite{fenton2009termination, luther2011low}.
In this sense, the quenching of excitation waves in fibrillation is considered in terms of \emph{ignition} -- the creation of new excitations; the electric field is applied to the whole tissue, and is only effective for the repolarized regions -- a relative minority of the overall tissue during the initial phase of defibrillation.
Developing techniques which can accurately predict the outcome of direct intervention into the propagation of excitation waves -- \emph{quenching} -- may reveal potential for improving the efficacy of defibrillation techniques by affecting the \emph{depolarized} regions of the tissue.
The brute-determination of quenching for excitation waves is untenable, and thus this study aims to determine whether similar semi-analytical linear techniques to those employed in the prediction of ignition for excitable media may be used for the prediction of quenching.}

\CM{If cardiac arrhythmia is understood as the presence of undesirable excitations in the heart muscle}, then the development of defibrillation techniques are targeted investigations of quenching in a specific context (i.e. cardiac arrhythmia with uncertain states).
The literature covering suppression of \CM{electrical} excitation waves \CM{in cardiac tissue} is vast, c.f. \CM{Ref.~\cite{dosdall2010mechanisms}} and citations therein.
Numerous methods are presented in the mathematical \CM{--} rather than medical \CM{--} literature, as well\CM{~\cite{fenton2009termination,buran2017control,detal2022terminating}, focusing on the transience of the fibrillating state}.
The suppression of \CM{\emph{stable}} traveling waves in \CM{low-dimensional} settings is less well-studied.
\CM{Ref.~\cite{osipov1999using}} investigated the quenching (`successful suppression') of stable pulses in slow-fast excitable media through the application of time-dependent control to the slow variable; they formalize this approach as the imposition of a lag on the wavefront such that the distance between the wavefront and the waveback of the excitation shrinks to zero in finite time.
This approach makes intuitive sense, and reflects one of the dominant pathways for wave break observed in models of cardiac electrical excitation~\CM{\cite{marcotte2017dynamical}}.
However, this approach \CM{relies on the application of control to components of the state which are typical inaccessible -- i.e. the ionic gates of the cells, rather than the potential across the cellular membrane --} and may over-suppress for models with realistic dissipation of waves expected to be relevant for cardiac electrical excitations~\CM{\cite{biktashev2002dissipation,rappel2022physics}}.
\CM{Semi-analytic techniques which incorporate details of the underlying physics typically improve efficiency over general numerical approaches by specializing to a particular mechanism~\cite{bezekci2017fast}.}

Phenomenological and detailed ionic models of excitable media possess, at minimum, a stable rest state $\bar{\vec{u}}$ representing the quiescent state to which unstimulated tissue relaxes.
Given the application of a sufficiently large (in amplitude and extent) perturbation to that rest state, the medium locally excites and extends the excited region by recruiting the energy stored in the quiescent excitable medium ahead of the front.
\CM{In mathematical models, these} fronts may persist indefinitely, or the medium may recover in finite time so that the front develops into a localized pulse which \emph{transiently} recruits the energy of the medium.
Front or pulse solutions are center-stable traveling wave solutions of the underlying \CM{partial differential equation (PDE)} model of the medium dynamics, and the minimal perturbation which stably transitions the system from the rest state to the excited state is the critical ignition perturbation.
Likewise, the minimal perturbation which stably transitions the system from the excited state to the rest state is the critical \emph{quenching} perturbation.

Throughout this \CM{study}, models of excitable media will take the form of a reaction-diffusion \CM{PDE},
\begin{equation}\label{eq:ivp}
	\partial_t\vec{u} - \bm{\mathrm{D}}\partial_x^2\vec{u} = \vec{f}(\vec{u}),
\end{equation}
where $\vec{u} = [u_1,\dots,u_m](t,x)$ and $\vec{f} = [f_1,\dots,f_m]$.
\CM{The reaction term $\vec{f}$ encodes the change in the state of a cell in terms of the electrical potential across the cellular membrane -- the transmembrane potential $u_1$ -- and the activation and inactivation of the ionic gates in the cell $u_2, \dots, u_m$ restricting or enabling the flow of ionic currents, e.g. sodium and potassium ions.}
The diffusion \CM{term is modulated by $\bm{\mathrm{D}} = \diag{1,0,\dots}$ to restrict diffusion to the transmembrane potential -- the fast component -- } $u_1$, requiring that the remaining components are non-diffusive.
\CM{By contrast, the slow components $u_2, \dots, u_m$ are typically inhibitory, and represent the ability of a cell to be excited -- refractoriness.}

The excitable system has a unique spatially uniform rest state denoted $\bar{\vec{u}}$, which satisfies $\vec{f}(\bar{\vec{u}}) = \vec{0}$, from which the ignition of excitations has been investigated previously~\CM{\cite{starmer1993vulnerability, biktashev2004non, starmer2007initiation, idris2007critical, biktashev2008initiation, idris2008analytical, bezekci2017fast, bezekci2020strength, marcotte2020predicting}}.
\CM{The uniform rest state is said to be \emph{quiescent} and corresponds to a state in which the tissue or cell is able to be excited; driving fibrillating tissue toward this state is desirable, as it permits the reassertion of the normal rhythm~\cite{dosdall2010mechanisms}.}
In a frame moving with speed $c$, \CM{equation}~\eqref{eq:ivp} becomes
\begin{equation}\label{eq:bvp}
	\vec{0} = \bm{\mathrm{D}}\vec{u}^{\prime\prime} + c\vec{u}^\prime + \vec{f}(\vec{u}),
\end{equation}
where $\vec{u}^\prime = \dd\vec{u}/\dd\xi$, and $\xi=x-ct$.
For sufficiently large domains, two solutions of \eqref{eq:bvp} persist for $c>0$, with the slower ($c=\hat{c} < \check{c}$) corresponding to an unstable traveling wave solution ($\hat{\vec{u}}$) and the faster ($c=\check{c} > \hat{c}$) corresponding to a stable traveling wave solution ($\check{\vec{u}}$)~\CM{\cite{simitev2011asymptotics}.}
When the slow wave has a single unstable mode, the center-stable manifold separates the stable rest state from the stable wave configuration in the whole state space.

\CM{Linearizing about the stable wave solution provides minimal dynamical information -- the leading mode corresponds to the equivariance of the state, and the rest are contracting.}
Linearizing about \CM{the \emph{unstable}} wave solution defines the comoving-frame operator \CM{$\op{L}(\hat{\vec{u}}, \hat{c})$} \CM{with left and right eigenmodes satisfying},
\begin{equation} \label{eq:evp}
	\hat{\vec{v}}_i \sigma_i =  \op{L}(\hat{\vec{u}},\hat{c}) \hat{\vec{v}}_i, \quad
	\hat{\vec{w}}_j^\dagger \sigma_j^* = \op{L}^\dagger(\hat{\vec{u}},\hat{c}) \hat{\vec{w}}_j^\dagger,
\end{equation}
such that the left eigenfunctions ($\hat{\vec{w}}_i$) are the unique projectors of the right eigenfunctions ($\hat{\vec{v}}_i$).
Though the solutions of \eqref{eq:evp} are formally scale-free, the left and right sets of eigenfunctions satisfy a \CM{bi-orthogonality} condition \CM{defined by the} inner product, $\delta_{ij} = \langle \hat{\vec{w}}_j | \hat{\vec{v}}_i \rangle = \int_{-\infty}^{+\infty} \dd \xi \, \hat{\vec{w}}_j^\dagger(\xi) \hat{\vec{v}}_i(\xi)$.
The eigenvalues, $\sigma_i$, determine the growth of each eigenmode $\hat{\vec{v}}_i$ over time.
\CM{Generally}, $\op{L}(\hat{\vec{u}}, \hat{c})$ is not self-adjoint, and the left and right eigenfunctions are distinct.
\CM{For the unstable pulse solution, $\sigma_1 > 0$ while $\sigma_2 = 0$, the latter corresponding to the translational equivariance of the unstable pulse.
While other choices for the basis describing the linear dynamics are possible, the eigenbasis has unique utility as it diagonalizes the operator and the leading (right) modes possess simple interpretations (instability, symmetry).}

\begin{figure}
	\includegraphics{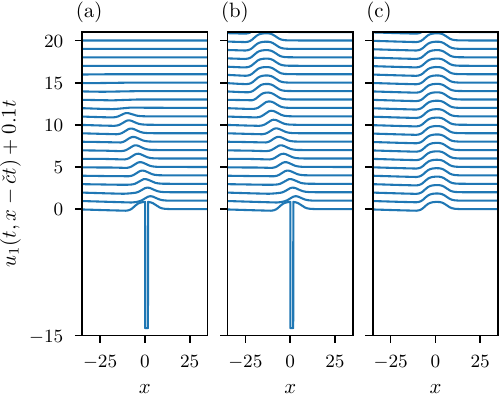}
	\caption{(a) Supercritical, (b) subcritical, and (c) unperturbed dynamics for the FitzHugh-Nagumo model \CM{stable pulse} in \CM{a co-moving frame} with speed $\check{c}$.
	For (a,b), the quenching perturbation is a rectangular envelope in the $u_1$ channel, centered at $x=1$, with width substantially smaller than the pulse \CM{($\sim5\%$)}, and amplitudes $\CM{U_{q}^\pm} = -18.8314 \mp 10^{-4}$ for the super- and sub-critical perturbations, respectively.}
	\label{fig:1}
\end{figure}

\CM{As the stable pulse is linearly stable, any quenching is due to finite-size perturbations to the state; for sufficiently small amplitudes, these perturbations are effectively linear and thus ineffective, while for larger amplitudes the perturbations quench the pulse.}
Figure~\ref{fig:1}(a) demonstrates quenching \CM{by direct numerical simulation (DNS)}; for super-critical amplitude perturbations to the stable excitation wave, the state approaches quiescence, $\lim_{t\to\infty}\vec{u}(t,x)\to \bar{\vec{u}}$.
For sub-critical amplitude perturbations the stable excitation wave recovers, shown in figure~\ref{fig:1}(b).
The sub-critical perturbation has the net effect of introducing some \CM{net displacement} $\delta > 0$ in the progression of \CM{the} wave relative to the unperturbed stable excitation wave, c.f., figure~\ref{fig:1}(b,c), $\lim_{t\to\infty}\vec{u}(t,x) \to \check{\vec{u}}(x-\check{c}t \CM{+ \delta)}$.
\CM{The essence of the quenching prediction is the determination of the critical quenching amplitude for a given perturbation envelope.}

\section{Theory}

In this \CM{S}ection we develop the theoretical groundwork for \CM{theory of quenching} used in this \CM{study}.
We first describe the essential ingredients of the classical linear ignition theory, and unify the methods for selecting a representative from the translational symmetry group using heuristics motivated by notions of distance in infinite-dimensional spaces.
We then extend the ignition theory to non-uniform stable states and detail the computation of the critical \CM{quenching} amplitude in this new context.
The final portion of this Section derives and simplifies the analogous heuristics from the linear ignition theory for the quenching problem, and develops a uniqueness argument based on the asymptotic properties of the perturbation.

\subsection{Ignition}

The classical ignition problem~\CM{\cite{starmer1993vulnerability, biktashev2004non, starmer2007initiation, idris2007critical, biktashev2008initiation, idris2008analytical, bezekci2017fast, bezekci2020strength}} considers the critical perturbation to the quiescent state for which the dynamics asymptotically approaches the stable excitation wave.
Identifying the initial condition as a perturbation from the rest state \CM{$\vec{u}(0,x) = \bar{\vec{u}} + \CM{\bar{\vec{h}}(x)}$, and equating} this configuration to a perturbed reference state selected from the group orbit of the unstable wave, $\hat{\vec{u}}(\xi) + \hat{\vec{h}}(\xi)$, where $\xi = x-s$ and $s$ an appropriate shift,
\begin{equation}
	\vec{u}(0,x) = \hat{\vec{u}}(\xi) + \hat{\vec{h}}(\xi) = \bar{\vec{u}} + \CM{\bar{\vec{h}}(x)},
\end{equation}
allows us to use information about the exact solution $\hat{\vec{u}}(\xi)$ to predict the long-term dynamics of \CM{$\vec{u}(0,x)$}.
\CM{We may predict the minimal amplitude of a localized perturbation to the rest state which successfully ``ignites'' the medium by taking an appropriate inner product with $\hat{\vec{w}}_1(\xi)$ and requiring the excitation of the unstable mode to vanish.}
As the rest state is uniform and invariant with respect to translations, and the unstable wave is equivariant under translations, \CM{for a prescribed perturbation envelope $\bar{\vec{X}}(x)$ the amplitude of the perturbation to the rest state may only be determined up to the the reference shift $s$, so that $\bar{\vec{h}}(x) = \bar{U}(s) \bar{\vec{X}}(x)$.}
Rearranging gives a compact equation for the critical ignition amplitude \CM{$\bar{U}(s)$} for a prescribed perturbation envelope \CM{$\bar{\vec{X}}(x)$},
\begin{equation}\label{eq:ignition}
	\bar{U}(s) = \frac{ \langle \hat{\vec{w}}_1(\xi) | \hat{\vec{u}}(\xi) - \bar{\vec{u}} \rangle }{ \langle \hat{\vec{w}}_1(\xi) | \bar{\vec{X}}(\xi+s) \rangle },
\end{equation}
where the appropriate shift $s$ is determined by a constraint equation, $0 = \langle \bar{\vec{\Phi}}_l(\xi) | \bar{\vec{X}}(\xi+s) \rangle$.
The constraint functional $\bar{\vec{\Phi}}_l(\xi)$ derives from heuristic considerations for the optimal reference frame\CM{, by extremizing a heuristic for the projective distance in the tangent space of the unstable pulse}, which we will \CM{consider again in} the application to quenching.
We include the frame selectors for the classical ignition problem for completeness,
\begin{align*}
	\bar{\vec{\Phi}}_1(\xi) &= \langle \hat{\vec{w}}_1 | \hat{\vec{u}} - \bar{\vec{u}} \rangle \hat{\vec{w}}_1^\prime(\xi), \\
	\bar{\vec{\Phi}}_2(\xi) &= \langle \hat{\vec{w}}_1 | \hat{\vec{u}} - \bar{\vec{u}} \rangle \hat{\vec{v}}_2(\xi) - \langle \hat{\vec{v}}_2| \hat{\vec{u}} - \bar{\vec{u}} \rangle \hat{\vec{w}}_1(\xi), \\
	\bar{\vec{\Phi}}_3(\xi) &= \langle \hat{\vec{w}}_1| \hat{\vec{u}}- \bar{\vec{u}} \rangle \hat{\vec{w}}_2(\xi) - \langle \hat{\vec{w}}_2 | \hat{\vec{u}} - \bar{\vec{u}} \rangle \hat{\vec{w}}_1(\xi),
\end{align*}
noting that each $\bar{\vec{\Phi}}_l(\xi)$ is formed explicitly from the available rest-state, unstable wave, and the leading left and right eigenfunctions of the unstable wave.
\CM{The frame selectors provide competing methods to determine an optimal reference shift $s$ for the ignition problem, by measuring projective distance in the tangent space of the unstable pulse solution.}
The constraint equation may be expressed as a function of the shift alone, $\bar{\mu}_l(s) = \langle \bar{\vec{\Phi}}_l(\xi) | \bar{\vec{X}}(\xi+s) \rangle$, where the inner product is evaluated over all $x$ by the cross-correlation integral of two vector functions with an implicit sum over the variable indices.
The critical excitation frame is determined by an appropriate root, $\bar{\mu}_l(s^*) = 0$, which can be assumed unique based on the slow-fast scaling of the left and right eigenfunctions~\cite{marcotte2020predicting}\CM{, which uniquely determines the value of \eqref{eq:ignition}}.
\CM{We adapt the critical ignition argument to quenching in the next section}.

\subsection{Quenching}

Given a stable \CM{pulse} solution, $\check{\vec{u}}(x-\check{c}t)$, we can consider an initial configuration \CM{$\vec{u}(0,x)$} expressed as a perturbation to the stable \CM{pulse},
\begin{equation}\label{eq:quenching}
	\CM{\vec{u}(0,x)} = \check{\vec{u}}(x) + \check{\vec{h}}(x-\theta; x_s),
\end{equation}
will relax to the rest state \CM{(i.e. $\lim_{t\to\infty}\vec{u}(t,x) \to \bar{\vec{u}}$) for a prescribed perturbation $\check{\vec{h}}(x-\theta; x_s)$ centered at $x=\theta$ with width $x_s$.}
\CM{Much like the linear theory for ignition, effective predictions of quenching require the selection of a reference frame for the unstable pulse.}
\CM{However, when considering $\check{\vec{u}}(x)$ in place of $\bar{\vec{u}}$, the construction of frame-determining functions requires additionally considering the spatial variation of the state.
When the state is uniform, the perturbation may always be arbitrarily shifted to the same coordinate frame as the state; when the state is non-uniform, the perturbation has an induced parametric shift in frame, denoted $\theta$, which describes the origin of the perturbation.}

The linear theory considers the linearization about the unstable solution, selected from along its group orbit, $\exp(+t\hat{c}\partial_x)\hat{\vec{u}}(x-\hat{c}t-s) = \hat{\vec{u}}(x-s)$ for all times $t \geq 0$.
We rewrite the initial configuration as the unstable solution plus a perturbation,
\[
	\vec{u}(t,x) = \hat{\vec{u}}(x-s) + \check{\vec{u}}(x) + \check{\vec{h}}(x-\theta; x_s) - \hat{\vec{u}}(x-s),
\]
which identifies $\hat{\vec{h}}(x-s) = \check{\vec{u}}(x) + \check{\vec{h}}(x-\theta; x_s) - \hat{\vec{u}}(x-s)$.
The construction of the theory is such that markedly different slow and fast waves limit the accuracy of the perturbation argument.
Linearizing about $\hat{\vec{u}}(x-s)$ and computing the excitation of the leading eigenmode,
\[
	\langle \hat{\vec{w}}_1(x-s) | \check{\vec{u}}(x) - \hat{\vec{u}}(x-s) + \check{\vec{h}}(x-\theta; x_s)  \rangle,
\]
\CM{which we require to vanish at time $t=0$}, to not excite the unstable mode, i.e. criticality.
Rearranging, we find
\[
	\langle \hat{\vec{w}}_1(x-s) | \check{\vec{h}}(x-\theta; x_s) \rangle = \langle \hat{\vec{w}}_1(x-s) | \hat{\vec{u}}(x-s) - \check{\vec{u}}(x) \rangle,
\]
\CM{i.e. that the negative projected deviation of the stable pulse from the reference pulse onto the leading mode is precisely the same as the perturbation to the stable pulse, itself.}
Defining $\check{\vec{h}}(x-\theta; x_s) = \CM{\check{U}}(s;\theta, x_s) \check{\vec{X}}(x-\theta; x_s)$ gives a parametric expression for the critical amplitude of the perturbation in $s$ and $\theta$,
\begin{equation}\label{eq:U}
	\check{U}(s; \theta, x_s) = \frac{ \langle \hat{\vec{w}}_1(\xi) | \hat{\vec{u}}(\xi) - \check{\vec{u}}(\xi + s) \rangle }{ \langle \hat{\vec{w}}_1(\xi) | \check{\vec{X}}(\xi+s-\theta; x_s) \rangle }\CM{.}
\end{equation}
Compare \eqref{eq:U} with \eqref{eq:ignition}, and the explicit coordinate parameter $\theta$ \CM{is immediately identifiable} as a relative phase for the perturbation along the stable wave.
\CM{Equation~\ref{eq:U}} is a function of the reference frame shift $s$, and parameterized by the perturbation envelope center $\theta$ and width $x_s$.

We now apply each heuristic to derive the quenching problem shift selectors, $\check{\vec{\Phi}}_l(\xi, s)$, which define the reference frame as the root of a parameterized inner product with the perturbation, i.e. $\check{\mu}_l(s^*) = 0$.
In the ignition case, the first heuristic minimizes the amplitude of the perturbation to $\bar{\vec{u}}$ over all frames $s$ by extremizing the numerator of \eqref{eq:ignition}\CM{, which} simplifies to $0 = \langle \hat{\vec{w}}_1^\prime(x-s) | \bar{\vec{h}}(x-s) \rangle$\CM{, as  $\partial_x\bar{\vec{u}} = \vec{0}$}.
For the quenching problem, $\partial_x\check{\vec{u}}(x) \neq \vec{0}$, the \CM{extremization} of \CM{\eqref{eq:U}} with respect to $s$ is determined by $0 = \partial_s \CM{\check{U}}(s;\theta, x_s)$, whose numerator expands to,
\begin{equation*}
\begin{split}
	0 &= \langle \hat{\vec{w}}_1(x-s) | \hat{\vec{u}}(x-s) - \check{\vec{u}}(x) \rangle \langle \hat{\vec{w}}_1^\prime(x-s) | \check{\vec{h}}(x-\theta; x_s) \rangle \\
		&- \langle \hat{\vec{w}}_1(x-s) | \check{\vec{h}}(x-\theta; x_s) \rangle \left( \langle \hat{\vec{w}}_1^\prime(x-s) | \hat{\vec{u}}(x-s) - \check{\vec{u}}(x) \rangle \right. \\ &\left. + \langle \hat{\vec{w}}_1(x-s) | \hat{\vec{u}}^\prime(x-s) \rangle \right).
  \end{split}
\end{equation*}
Rearranging to form $\check{\vec{\Phi}}_1(x-s; s)$,
\begin{align*}
	\check{\vec{\Phi}}_1(x-s; s) =& \, \langle \hat{\vec{w}}_1(x-s) | \hat{\vec{u}}(x-s) - \check{\vec{u}}(x) \rangle \hat{\vec{w}}_1^\prime(x-s) \\
	&- \langle \hat{\vec{w}}_1^\prime(x-s) | \hat{\vec{u}}(x-s)  - \check{\vec{u}}(x) \rangle\hat{\vec{w}}_1(x-s),
\end{align*}
where we have used $\langle \hat{\vec{w}}_1(x-s) | \hat{\vec{u}}^\prime(x-s) \rangle = \langle  \hat{\vec{w}}_1(x-s) | \hat{\vec{v}}_2(x-s) \rangle = 0$, guaranteed by the biorthogonality of the eigenmodes.

The second heuristic minimizes the $L_2$-distance between the perturbed solution and the unstable reference solution over all choices of frame parameterized by the shift $s$,
\[
	s_2 = \arg\min_s \, \langle \vec{u}(x) - \hat{\vec{u}}(x-s) | \vec{u}(x) - \hat{\vec{u}}(x-s) \rangle,
\]
for which a necessary condition is that the $L^2$ norm of $\vec{u}(x) - \hat{\vec{u}}(x-s)$ is extremal, i.e. $\partial_s \langle \vec{u}(x) - \hat{\vec{u}}(x-s) \rangle^2 = 0$, which simplifies using $\hat{\vec{v}}_2(x-s) = -\partial_s\hat{\vec{u}}(x-s)$,
\[
	0 = \langle \hat{\vec{v}}_2(x-s) | \check{\vec{u}}(x) + \check{\vec{h}}(x-\theta; x_s) - \hat{\vec{u}}(x-s) \rangle,
\]
from which we rearrange to identify $\check{\vec{\Phi}}_2(x-s;s)$,
\begin{align*}
	\check{\vec{\Phi}}_2(x-s;s) =& \langle \hat{\vec{w}}_1(x-s) | \hat{\vec{u}}(x-s) - \check{\vec{u}}(x) \rangle \hat{\vec{v}}_2(x-s) \\
	&- \langle \hat{\vec{v}}_2(x-s) | \hat{\vec{u}}(x-s) - \check{\vec{u}}(x) \rangle \hat{\vec{w}}_1(x-s).
\end{align*}

The third heuristic requires that the perturbation be orthogonal to the space spanned by the first and second right eigenfunctions; as the first eigenmode appears in the constraint in the derivation of \eqref{eq:U}, this is satisfied by the projection onto the second left eigenmode, i.e. the translational response function,
\[
	0 = \langle \hat{\vec{w}}_2(x-s) | \check{\vec{u}}(x) + \check{\vec{h}}(x-\theta; x_s) - \hat{\vec{u}}(x-s) \rangle,
\]
which may be rearranged to form $\check{\vec{\Phi}}_3(x-s; s)$,
\begin{align*}
	\check{\vec{\Phi}}_3(x-s;s) =& \langle \hat{\vec{w}}_1(x-s) | \hat{\vec{u}}(x-s) - \check{\vec{u}}(x) \rangle \hat{\vec{w}}_2(x-s) \\
	&- \langle \hat{\vec{w}}_2(x-s) | \hat{\vec{u}}(x-s) - \check{\vec{u}}(x) \rangle \hat{\vec{w}}_1(x-s),
\end{align*}
which we identify as $\check{\vec{\Phi}}_2(x-s;s)$ where the second right eigenfunction ($\hat{\vec{v}}_2(\xi)$) is replaced with its projector ($\hat{\vec{w}}_2(\xi)$).

\CM{The frame-selectors deviate from the ignition problem formalism, as $\check{\vec{\Phi}}_l(x-s; s)$ transforms with $s$ as the product of vectors with $(x-s)$ dependence\CM{,} weighted by scalars with $s$ dependence,
\begin{equation}\label{eq:Phi_generic}
    \check{\vec{\Phi}}_l(x-s; s) = \phi_{l,1}(s) \vec{a}_{l,1}(x-s) + \phi_{l,2}(s) \vec{a}_{l,2}(x-s).
\end{equation}
This change in the transformation of $\check{\vec{\Phi}}_l(x-s; s)$ from the ignition formalism informs a new approach to frame selection, as the utility of the projector formalism is limited.
For simplicity}, we form the scalar functions
\begin{equation}\label{eq:mu}
	\check{\mu}_l(s; \theta, x_s) = \langle \check{\vec{\Phi}}_l(x-s;s) | \check{\vec{X}}(x-\theta; x_s) \rangle,
\end{equation}
which reduces over the spatial coordinate $x$ and the variable index, and whose root(s) define potential reference frame(s) for a particular perturbation \CM{envelope $\check{\vec{X}}(x-\theta;x_s)$}.
The question of identifying a \emph{unique} frame amounts to selecting a \emph{single} root of \eqref{eq:mu} \CM{for each heuristic.}

Indeed, we have no strong guarantees for the uniqueness of the roots, for the general case.
We may, however, motivate a procedure which guarantees a unique root for all \CM{viable} $x_s$ provided a unique root exists for $x_s \to \infty$.
We assume the perturbation envelope is parameterized such that \CM{for sufficiently large $x_s$,} it asymptotically approaches a constant vector of length $m$ with unit $L^\infty$-norm,
\[
    \lim_{x_s\to\infty} \|\check{\vec{X}}(x-\theta;x_s)\|_\infty \to 1,
\]
implying \CM{$\partial_\theta \check{U}(s; \theta, x_s) = \partial_{x_s} \check{U}(s; \theta, x_s) = 0$} in the limit.
For simplicity, we assume that the asymptotically wide perturbation is entirely in the $u_1$-channel of the state, i.e. $\lim_{x_s\to\infty}\check{\vec{X}}(x-\theta;x_s) \to [1,0,\dots]$.
\CM{This comports with the typical interpretation of $u_1$ as the transmembrane potential, and thus observable (with micro-electrode measurements) and controllable (with current stimulus).}
In the asymptotic\CM{ally wide} regime, we require a means of identifying a unique root $s^*$ of \eqref{eq:mu}, which we take to be the most strongly negative option of \eqref{eq:U}.
For $\lim_{x_s\to\infty} \CM{\check{U}}(s; \theta, x_s) < 0$, we \CM{rearrange \eqref{eq:U} and} require,
\[
    \langle \hat{\vec{w}}_1(\xi) | \hat{\vec{u}}(\xi) - \check{\vec{u}}(\xi+s) \rangle \langle \hat{\vec{w}}_1(\xi) | \check{\vec{X}}(\xi+s-\theta;x_s) \rangle < 0.
\]
The second factor can be guaranteed positive due to the biorthogonality of the eigenfunctions, and the first may be asserted negative based on the construction of the slow and fast waves and the triangle inequality; \CM{therefore} we may select a shift $s$ which extremizes $\langle \hat{\vec{w}}_1(\xi) | \check{\vec{u}}(\xi+s) \rangle$ while leaving $\langle \hat{\vec{w}}_1(\xi) | \hat{\vec{u}}(\xi) \rangle$ constant.
Given the asymptotic solution $\lim_{x_s\to\infty} \CM{\check{U}}(s; \theta, x_s)$, we may trace the branch of solutions to smaller perturbation widths until the solution vanishes, and thus construct a unique prediction for the critical quenching amplitude \CM{for all viable $x_s$} through continuation.

We have published example code and data for the linear theory prediction of critical quenching perturbations online~\cite{gitrepo}.
This code only requires the location of files containing $\bar{\vec{u}}$, $\check{\vec{u}}$, $\hat{\vec{u}}$, and $\hat{\vec{v}}_1$, $\hat{\vec{v}}_2$, $\hat{\vec{w}}_1$, and $\hat{\vec{w}}_2$, on a consistent domain.
The code performs estimation of the error of the method by comparing the construction of $\check{\mu}_l(s; \theta, x_s)$ \CM{using spectral cross-correlation} against $\langle \check{\vec{\Phi}}_l(x-s;s) | \check{\vec{X}}(x-\theta)\rangle$, where \CM{the latter} is computed by explicit translation of the vector components.

\section{Methods}

The essential ingredients of the application of the extended linear theory to a prescribed perturbation envelope $\check{\vec{X}}(x-\theta; x_s)$, are the stable solution $\check{\vec{u}}$, the unstable solution, $\hat{\vec{u}}$, the leading two left eigenfunctions $\hat{\vec{w}}_1$ and $\hat{\vec{w}}_2$, and their associated derivatives.
The solutions themselves are computed approximately using continuation within the \textsc{Auto-07p}~\cite{doedel2007auto} framework.
The solutions are then refined using a global spectral expansion, and the refined solutions used as non-constant-coefficient fields in the forward and adjoint eigenproblems \CM{\eqref{eq:evp}, using} the \textsc{Dedalus2}~\cite{burns2020dedalus} framework.
From these quantities, any of the $\check{\mu}_l(s; \theta, x_s)$ may be formed by computing cross-correlation integrals, likewise the prediction of $\Us$ using the formalism laid out in in the previous section.
\CM{The continuation of a root $s^*$ for large $x_s$ to small $x_s$ is done using so-called `natural continuation', and thus avoids the complexity of bifurcations of the roots at the expense of under-resolving regions where the dependence with $x_s$ is significant.}

To verify the linear theory predictions, we form initial conditions parameterized by perturbation amplitude \CM{$U_q < 0$, and perturbation envelope $\check{\vec{X}}(x-\theta;x_s)$} width $x_s$ and position $\theta$,
\[
	\CM{\vec{u}(0,x)} = \check{\vec{u}}(x) + \CM{U_q}\check{\vec{X}}(x-\theta; x_s),
\]
\CM{and solve the resulting initial-value problem using direct numerical simulation.}
For the purposes of testing the predictions of the linear theory, the perturbation is defined by a top-hat function centered at $\theta$,
\begin{equation}\label{eq:pert}
	\check{\vec{X}}(y; x_s) = [1,0,\dots] \, H(y+x_s/2) \, H(x_s/2-y),
\end{equation}
where where $y = x-\theta$ and $H(z) = (1+\mathrm{sign}(z))/2$ approximates the Heaviside distribution.
The shape is defined such that $\|\check{\vec{X}}_1(x-\theta; x_s)\|_{\infty} = 1$ and $\|\check{\vec{X}}_1(x-\theta; x_s)\|_{1} = x_s$, so that the perturbation is \CM{restricted to the voltage channel, or fast variable of the excitable model}.
\CM{T}his choice of perturbation \CM{envelope} is chosen for its simplicity; the methods described in this \CM{study} are not specific to this \CM{envelope or parameterization}.

We solve the initial-boundary-value problem \eqref{eq:ivp} with initial condition \CM{$\vec{u}(0,x)$} and periodic boundary conditions over $x \in [0,L)$ and a fixed time interval $t \in [0, T]$ with $T = L/(2\check{c})$, using the \textsc{Dedalus2} framework initial-value problem (IVP) solver.~\footnote{Reaction-diffusion models with pulse solutions may have a ``one-dimensional spiral wave''~\CM{\cite{cytrynbaum2009global}} solution, which consists of repeated back-ignition and eventual self-interaction due to propagation around the periodic domain. Therefore two stable pulses traveling in opposite directions from the same original position can not interact destructively sooner than a significant proportion of $t > T$.}
We simultaneously track the state over time, $\vec{u}(t,x)$, and compare the final state, using an appropriate measure, to the exact solutions.
This procedure amounts to the solution of a one-dimensional root-finding problem for each width and shift pair for the critical quenching amplitude.

We track the development of the perturbed stable wave over time by computing the absolute-valued deviation of the fast variable from the \CM{corresponding rest value},
\begin{equation}\label{eq:psi}
	\psi(t) = \int_0^L \dd x \, |u_1(t,x) - \bar{u}_1|,
\end{equation}
and compare the final value $\psi(T)$ to the respective evaluation at $\bar{u}_1$, $\hat{u}_1$, and $\check{u}_1$, which we designate by $\bar{\psi} \equiv 0$, $\hat{\psi}$, and $\check{\psi}$, respectively.
\CM{The form of the diagnostic function \eqref{eq:psi} was chosen to test whether it was necessary to used the same norm as the definition of the left eigenfunctions \eqref{eq:evp}, and whether partial state information precludes accurate determination of the state.}
The critical amplitude then is defined by the root of the function,
\begin{equation}\label{eq:rootfinding}
	\varphi(U_q) = \lim_{t\to\infty}\psi(t; U_q) - \hat{\psi},
\end{equation}
\CM{and} $\lim_{t\to\infty}\psi(t; U_q) \in \{\bar{\psi}, \hat{\psi}, n\check{\psi} - (n-1)\check{\delta}\}$ where the last option accounts for the possibility of transient back-ignition ($n \in \mathbb{N}$ denotes the number of stable pulses) and $\check{\delta}$ is the overlap deformation correction to $\psi(T)$ from the presence of multiple stable pulses on the same domain.
As \CM{the integration interval} $T\to\infty$, the probabilities of \CM{$\psi(T; U_q)$} taking on these values approaches $\{ P_q, 0, 1-P_q\}$, where $P_q$ is the probability that $U_q$ is quenching for a provided $(x_s,\theta)$ pair.
While no diagnostic function is able to uniquely determine the ultimate attractor of any arbitrarily configured state, in a large neighborhood centered on each of the exact stable solutions, this intuition holds.
We ensure that our final states are in such a neighborhood by solving the IVP over a sufficiently long time interval.

\begin{figure}
\includegraphics[scale=1]{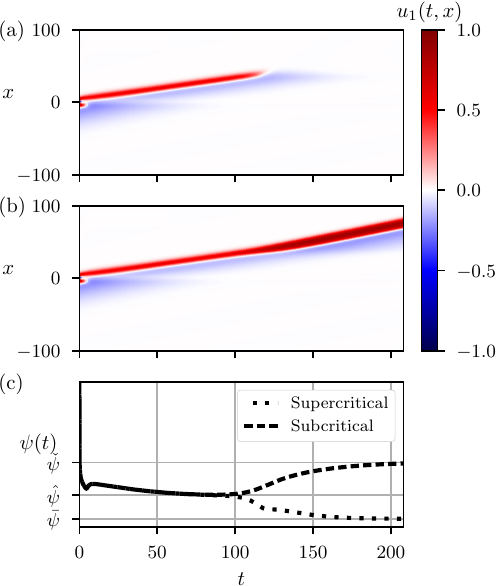}
\caption{\CM{Space-time dynamics of the fast variable $u_1(t,x)$ for} (a) Super-critical and (b) sub-critical \CM{perturbations to the stable pulse} of the FitzHugh-Nagumo model \eqref{eq:fhn} corresponding to figure~\ref{fig:1}(a,b), and (c) their diagnostic function $\psi(t)$ for the (dots) super-critical and (dash) sub-critical dynamics.}
\label{fig:2}
\end{figure}

Figure~\ref{fig:2}(a,b) depicts the super- and sub-critical \CM{space-time} dynamics for the quenching perturbations along with (c) their diagnostic function $\psi(t)$.
While $\psi(0)$ is very close between the two solutions, $|\psi_{-}(0) - \psi_{+}(0)|\approx |(U_{q}^{-}-U_{q}^{+}) x_s| = 3 \times 10^{-4}$, they diverge after $t\approx 100$, which corresponds to the close passage of the state to the unstable pulse solution, \CM{$\hat{\vec{u}}(x-s')$}, and corresponding measure $\hat{\psi}$.
After $t\approx 200$, the two solutions have diverged and are very similar to the two stable states, $\bar{\vec{u}}$ and \CM{$\check{\vec{u}}(x-s'')$}, in (a,b), respectively.
Further refinement of the critical perturbation amplitude for this $(x_s,\theta)$ pair may be attained using a bisection root-finding algorithm, limited only by the \CM{error} of the time-stepping.

\section{Results}

In this Section, we first introduce an excitable model, and remark on its inclusion and relevance to the task at hand.
We proceed to describe the relevant linear theory ingredients and the efficacy of these ingredients in predicting the \CM{critical quenching amplitude}, across a range of perturbation widths and positions.
We finish the discussion of each model by comparing the \CM{direct numerical simulation (DNS)} results and linear theory predictions, and \CM{detail any} limitations of the latter for predicting the former.

\subsection{FitzHugh-Nagumo model}

\begin{figure}[htpb]
	\includegraphics{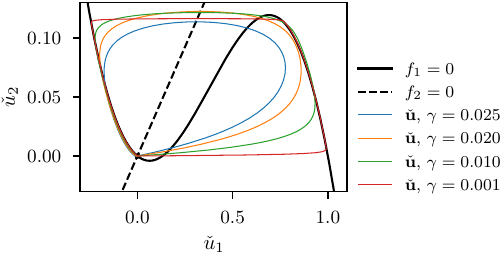}
	\caption{Nullclines of the FitzHugh-Nagumo model (black; $f_1=0$ solid, $f_2=0$ dashed) and the stable pulse solutions for $\gamma \in \{0.025, 0.020, 0.010, 0.001\}$.}
	\label{fig:3}
\end{figure}

\begin{figure}[htpb]
	\includegraphics{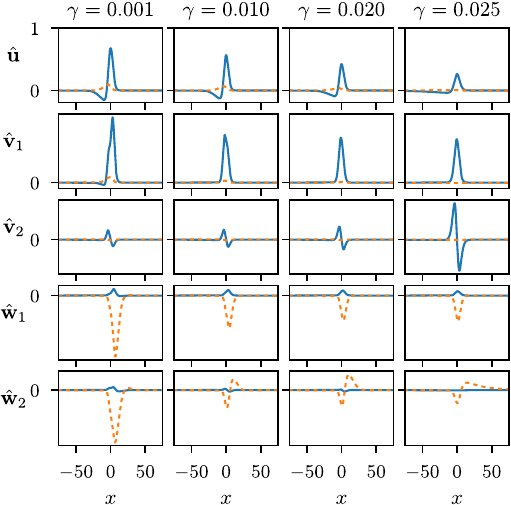}
	\caption{Linear theory ingredients for the FitzHugh-Nagumo model. (Top row) Unstable pulse solution $\hat{\vec{u}}$ for $\gamma \in \{0.001, 0.010, 0.020, 0.025\}$, with (second row) $\hat{\vec{v}}_1$, (third row) $\hat{\vec{v}}_2$, (fourth row) $\hat{\vec{w}}_1$, and (fifth row) $\hat{\vec{w}}_2$. \CM{The first component of each variable is the solid (blue) curve, and the second component is shown as dashed (orange).}}
	\label{fig:4}
\end{figure}

The \CM{FitzHugh-Nagumo model is archetypal of excitable media, and} arises from a simplification of squid axon signal dynamics~\cite{fitzhugh1961impulses}.
\CM{The FitzHugh-Nagumo model simplifies a Hodgkin-Huxley-type model, and represents the fast dynamics of the transmembrane potential in $u_1$ and the slow recovery dynamics of sodium and potassium channel deactivation in $u_2$~\cite{Sherwood2013}.}
The \CM{applicability} of the model \CM{to} cardiac \CM{excitation} is tenuous, but remains \CM{essential} for \CM{testing} new approaches to the study of the fundamental properties of excitation waves.
The model consists of two non-dimensional state variables interacting in a slow-fast system,
\begin{equation} \label{eq:fhn}
	f_1 = u_1(1-u_1)(u_1-\beta)-u_2, \quad f_2 = \gamma(\alpha u_1 - u_2),
\end{equation}
where we identify $u_1$ as the fast component and $u_2$ as the slow component, based on the time-scale separation $\gamma \ll 1$.
For the purposes of this study we fix $\alpha=0.37$ and $\beta=0.131655$, and vary $\gamma$, to consider morphologies of the fast and slow waves which are aligned with and counter to the underlying assumptions of the linear method.
We choose the time-scale separation $\gamma \in \{0.001, 0.010, 0.020, 0.025\}$ to explore differences in the morphology of the fast and slow pulses, c.f. fig.~\ref{fig:3} \& fig.~\ref{fig:4}, respectively.
In the limit of $\gamma\to 0$, the slow pulse speed vanishes, $\lim_{\gamma\to 0}\hat{c}\to 0$, forming a critical nucleus; meanwhile, in the same limit, the fast pulse loses post-excitation recovery, developing into a stable propagating front with finite speed $\lim_{\gamma\to 0} \check{c} > 0$.
\CM{As $\gamma$ increases, the fast and slow waves converge until coincidence for $\gamma_c \approx 0.026$.}

Figure~\ref{fig:3} depicts periodic stable pulse solution for each choice of $\gamma$ over the nullclines of the model.
Each stable solution is interpolated using $M=4096$ Chebyshev modes on a periodic domain of length $L=2700$ with dealiasing factor $N/M = 2$.
These solutions are resolved with projective boundary conditions and \CM{are} approximately homoclinic \CM{(to deviations smaller than $10^{-13}$)}.
To our knowledge there exists no generally applicable exact solution for the fast pulse on unbounded domains with non-vanishing slow-fast time-scale separation parameter~\cite{keener1980waves}, but appropriate asymptotics hold approximately~\cite{simitev2011asymptotics}.

The scaling of the critical pulse and the leading eigenmodes was investigated in a previous work~\cite{marcotte2020predicting}, which is most relevant for $\gamma\to 0$.
In the range $10^{-3} \leq \gamma < \gamma_c$, the scaling of the components of the pulse and eigenmodes change slowly, but the components of the pulse and eigenmodes change shape significantly as $\gamma$ varies.
The unstable solutions and their associated eigenfunctions are resolved \CM{with the same discretization and to the same maximal deviation as the stable pulses}.
The linear theory ingredients are depicted in figure~\ref{fig:4}.
The pulses are oriented such that the peak of the fast component occurs precisely at the origin: $0 = \arg\max \hat{u}_1(x)$, for all $\gamma$, matching the orientation of the fast pulse solutions (\CM{i.e.} $0 = \arg\max \check{u}_1(x)$).
The left and right eigenfunctions are depicted such that the appropriate pairs satisfy $\langle \hat{\vec{w}}_i | \hat{\vec{v}}_i \rangle = 1$ for $i=1,2$, and oriented such that $\hat{{v}}_1^{(1)}(0) > 0$ and $\partial_x\hat{{v}}_2^{(1)}(0) < 0$, for consistency \CM{with the interpretation as leading instability and translational modes, respectively}.

\begin{figure*}[hbtp]
	\includegraphics[width=\textwidth]{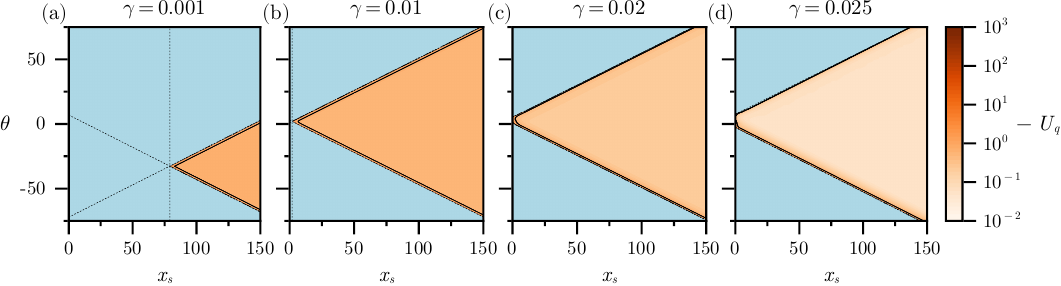}
	\caption{Critical quenching perturbations sampled over the $(x_s,\theta)$-plane, for (a) $\gamma=0.001$, (b) $\gamma=0.010$, (c) $\gamma=0.020$, and (d) $\gamma=0.025$.
		In all cases, the perturbation quenching amplitude $U_q$ determines the red color and blue denotes regions in which \eqref{eq:rootfinding} has no solution for all bounding samples (c.f. black dots in figure~\ref{fig:6}(a-d)).
		The $|\beta-1|$ asymptotic estimate contour is denoted by a solid black curve and the linear bounding region between successful and unsuccessful quenching searches is denoted by black dashed lines.}
	\label{fig:5}
\end{figure*}

We present direct numerical simulation results for a large number of realizations of the quenching problem in figure~\ref{fig:5} \CM{as an exploration of the bounds of quenching in terms of the perturbation parameters and verification of the spectral initial-value problem solver}.
For this exploration, \eqref{eq:ivp} with \eqref{eq:fhn} is solved on an interval of $0 \leq t \leq L/(2\check{c})$ with an $N=1+2^{13}$ point spatial discretization and $O(h^{12})$ centered-difference approximation of the diffusion operator, where $h = L/(N-1)$, using CVODE in the Sundials package~\cite{rackauckas2017differentialequations, gardner2022sundials, hindmarsh2005sundials} with a matrix-free GMRES solver for the linear system.
These numerical parameters reproduce the constant $\psi(t)$ for a stable pulse over the $T = L/(2\check{c})$ interval to a tolerance of $|\psi(T) - \psi(0)| < 10^{-5}$.
For these large finite-difference approximations, we focus on the root-finding problem over a bounded region of the $(x_s,\theta)$-plane.

Figure~\ref{fig:5} shows the \CM{DNS} results of the quenching problem for the stable pulses shown in figure~\ref{fig:3}.
Note that the perturbation width is positively-valued, and perturbations with unresolvable widths, i.e. $x_s < 2L/N$, are restricted.
The perturbation parameter $\theta$ is chosen to position the perturbation behind ($\theta < 0$) or ahead ($\theta > 0$) of the peak for the stable pulse.
The color designates the solution to \eqref{eq:rootfinding}: either a solution to the bisection problem is found (oranges) or \emph{no solution} is found (blue) for $U_q \in [-10^{3},0)$.
A narrow gap (white) between the successful and unsuccessful regions exists as the triangulation for both regions is conservative.
While it is possible that $U_q \to -\infty$ in the white region, we expect that actually the finite cutoff is real in the sense that the largest amplitude for a successful quenching is substantially smaller than the root-finding limit.
That is, if we expect that $U_q$ diverges for some $(x_s,\theta)$, then we should find saturation of $U_q \approx -10^{+3}$ near the bounds of the effective quenching regions.
We do not observe this in the data, as $|U_q| \ll 2\times 10^{+2}$, across the entire range of $\gamma$, and $|U_q| < 10^{+1}$ for $\gamma = 0.001$.

\begin{figure}[htpb]
	\includegraphics[width=\columnwidth]{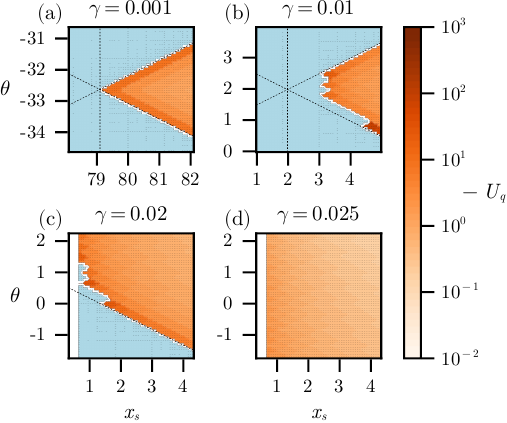}
	\caption{Critical quenching perturbations near the smallest quenching perturbation width for $\gamma \in \{0.001, 0.01, 0.02, 0.025\}$ in (a-d), respectively. Colors are as in figure~\ref{fig:5}(a-d), with black dots representing sampled quenching parameters.}
	\label{fig:6}
\end{figure}

The successful quenching parameters are bounded in the $(x_s,\theta)$-plane by lines with slope approximately $\pm 1/2$, which intersect at $(x_s^0,\theta^0)$ coordinates which correspond to the quenching perturbation with minimal width and corresponding optimal position.
We compute the slopes and intercepts using an exterior penalty method and find the deviation in the discriminator slopes from $\pm 1/2$ is $< 5 \times 10^{-3}$.
For sufficiently small time-scale separation parameter, $\gamma < 0.02$, we find that the minimal width is finite, corresponding to $x_s^0 \approx 79$ and $x_s^0 \approx 2$ for $\gamma=0.001$ and $\gamma=0.010$, respectively, cf. figure~\ref{fig:6}(a,b).
For $\gamma \geq 0.020$, we find that the linear bounding estimate for the minimum quenching width vanishes; i.e. the linear bounds of the successful quenching parameters intersect at $x_s^0 < 0$.
The successful quenching numerics fall short of resolving infinitesimally thin perturbations, bounded from below by the grid spacing $L/(N-1)\approx 0.33$, but they confirm this bound up to the discretization limit, cf. figure~\ref{fig:6}(c,d).
The reproduction of the limited region of quenching possibility is an important test for the predictive quenching theory.

\begin{figure*}[htpb]
	\includegraphics{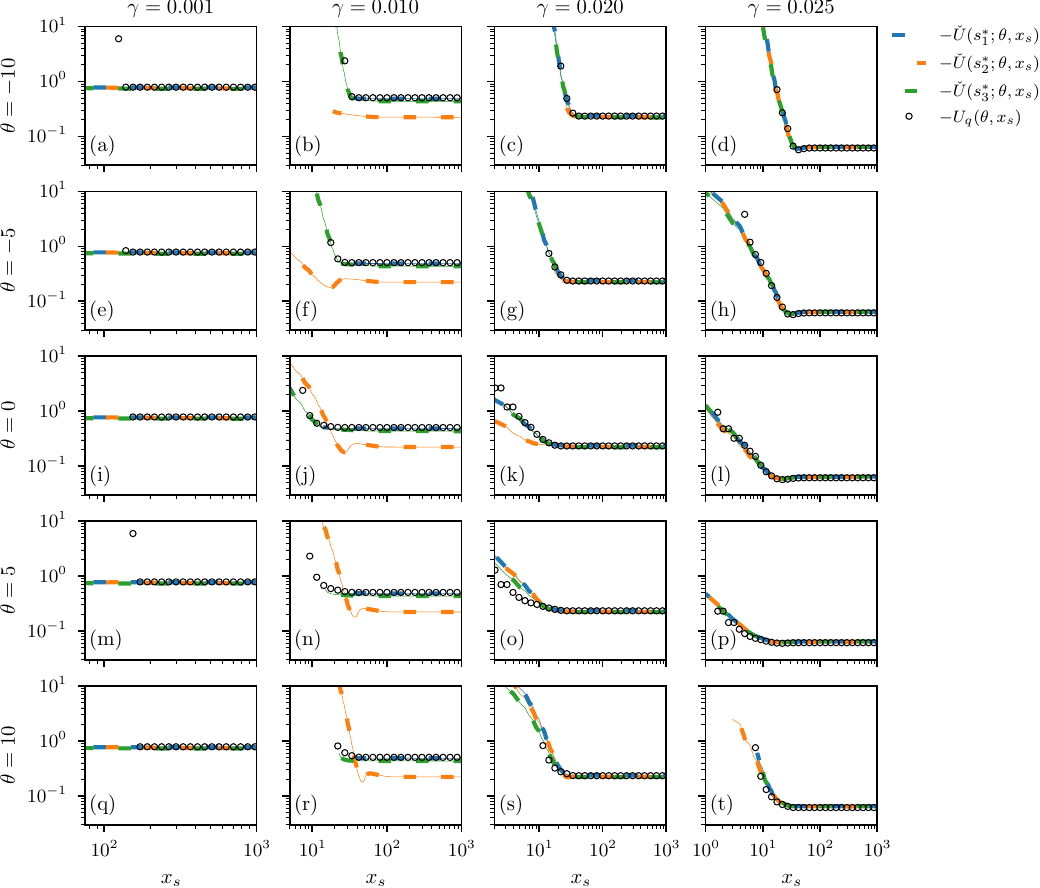}
	\caption{Critical quenching predictions for the FitzHugh-Nagumo model (color) using selectors \CM{$\vmu{1}$ (blue), $\vmu{2}$ (orange), and $\vmu{3}$ (green) compared to the} quenching computed using \CM{DNS} of the perturbed stable pulse (black, $\circ$).}
	\label{fig:7}
\end{figure*}

Predictions of the critical quenching amplitude (color dashes) and \CM{DNS} results (black $\circ$) are compared in figure~\ref{fig:7}, for $\theta \in \{-10,-5,0,5,10\}$, and for $x_s \in [2L/N, L/2]$.
The linear theory predictions agree with the \CM{DNS} results across one or more decades of $x_s$, for some values of $\theta$, for most choices of \CM{heuristic} $l$ and $\gamma$.
The prediction is most effective when the stable and unstable pulses are close in shape, as the $\gamma=0.025$ quenching amplitudes are reliably predicted across a wider range of $\theta$, $x_s$, and $U(s; \theta, x_s)$ than for $\gamma=0.020$, as explained in the theory construction.
For $\gamma=0.025$, perturbations centered at or immediately ahead of the peak of the stable pulse ($\theta \geq 0$) are able to quench the excitation for smaller perturbation widths and smaller amplitudes, while perturbations centered behind the peak of the pulse ($\theta < 0$) are successful over a smaller range of perturbation widths and the rate at which the critical quenching amplitude grows is significantly faster, c.f. (d) and (t).
This is expected---as a narrow perturbation appears at $\theta > 0$ and $t=0$, the stable pulse effectively runs into the perturbation by time $(\theta-x_s/2)/\check{c} \ll L/2\check{c}$, which drives the dynamics away from the stable solution.
Meanwhile, for narrow perturbations which are centered behind the peak of the wave, only the main excited region and tail of the stable pulse are affected---in this sense it is the front which controls the effectiveness of quenching perturbations.
The intuition that front recovery in FitzHugh-Nagumo models is robust thus appears correct, at least for sufficiently small $\gamma$ or large time-scale separation\CM{~\cite{biktashev2002dissipation}.}
\CM{However,} this important feature is not reflected in the predicted quenching amplitudes, which are approximately invariant with respect to perturbation width to much smaller $x_s$.

The separation between the slow and fast wave morphology impacts the asymptotic critical amplitude.
This is easiest to see for $\gamma=0.001$, where the steep wave front and wave back of the stable pulse lend itself to a rectilinear approximation.
The stable pulse $\check{u}_1(x)$ crosses zero at $x\approx x_a$ and $x\approx x_b$, and $0.8 \lesssim \check{u}_1(x) \lesssim 1$ in the excited region.
This indicates that for $\theta \approx (x_a+x_b)/2$ and $x_s \gtrsim |x_b-x_a|$, all quenching perturbations should succeed with $U(s;\theta, x_s)\gtrsim \beta-1$ (the maximal distance between the unstable and excited $f_1=0$ nullclines).
In figure~\ref{fig:7}(a,e,i,m,q), the critical quenching amplitude has $|U(s;\theta, x_s)| \approx 0.8$ which is comparable to $1-\beta \approx 0.86$.
Likewise, figure~\ref{fig:5}(a) shows there is little variation in the value of $|U_q|$ within the successful region, and that the region is approximately bounded by $|\beta-1|$.
Unexpectedly, the critical amplitude varies with the timescale separation parameter $\gamma$, as $\beta$ is held consistent throughout.
This indicates that our expectation from the asymptotic solution, $U_q \to \beta-1$, is not relevant for time-scale separations as small as $\gamma \gtrsim 0.01$.
\CM{The parameter value $\gamma = 0.01$ corresponds to a bifurcation for the stable rest state -- changing the relaxation dynamics from a stable node to a stable spiral -- however this bifurcation is not relevant for the quenching problem, as we consider perturbations to states only where they are far from the rest state.}

\subsection{Mitchell-Schaeffer model}

The Mitchell-Schaeffer model~\cite{mitchell2003two} is a simplified model of cardiac cell excitation, combining two state components to produce a realistic description of action potential shape and restitution.
It is derived from an asymptotic reduction of the Fenton-Karma model~\cite{fenton1998vortex} of atrial excitation by the adiabatic elimination of the fastest processes.~
\CM{The model is known to reproduce important features of cardiac action potential excitations from more complex cardiac models, and exhibit dynamical features relevant to cardiac modelling such as subcritical alternans.}
The Mitchell-Schaeffer model preserves multiple decay timescales,~\cite{alonso2016nonlinear} which we recombine to form an explicit time-scale separation and slow-fast system,
\begin{align}\label{eq:mitsch}
	f_1 &= u_1^2 (1-u_1)u_2 - u_1 \tau_i/\tau_u, \\
	f_2 &= \varepsilon \left( (1-u_2)H_k(u_g-u_1) \tau_c/\tau_o - u_2 H_k(\CM{u_1}-u_g)\right),\nonumber
\end{align}
where \CM{the fast variable $u_1$ again represents the potential across the cell membrane, the slow variable $u_2$ controls the activation and inactivation of ionic gates necessary for the functioning of the cell, and} we have introduced $H_k(x) = (1 + \tanh(k x))/2$.
The standard parameter values are $\tau_i=0.3~\mathrm{ms}$, $\tau_o=120~\mathrm{ms}$, $\tau_u=6~\mathrm{ms}$, $\tau_c=150~\mathrm{ms}$, $u_g=0.03$, and $k=100$.
We keep $\tau_i/\tau_u = 0.05$ and $\tau_c/\tau_o=1.25$ fixed, while allowing the non-dimensional ratio of the longest and shortest time-scales, $\varepsilon$, to vary.

\begin{figure}
    \centering
    \includegraphics{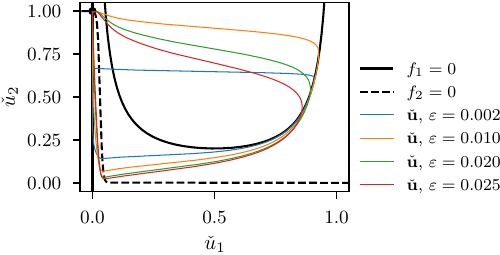}
    \caption{Nullclines of the Mitchell-Schaeffer model ( $f_1=0$ black, solid; $f_2 = 0$ black, dashed) and the stable pulse train solutions for $L=500$ for $\varepsilon\in\{0.002, 0.01,0.02, 0.025\}$.}
    \label{fig:8}
\end{figure}

Additionally, we have treated the Mitchell-Schaeffer model differently from the FitzHugh-Nagumo model and considered \emph{pulse train} solutions of the model instead of \emph{homoclinic} solutions.
This ultimately assesses the ability of the linear theory to predict quenching perturbations for pulses which are not isolated, i.e. in analogy to tachycardia-like scenarios.
Figure~\ref{fig:8} shows the nullclines of the model with the stable pulse train solutions on a ring of length $L=500$, for $\varepsilon\in\{0.002, 0.01,0.02, 0.025\}$.
We choose $\varepsilon \geq \tau_i/\tau_c$ to improve the stiffness of the model while permitting the study of quenching for fast solutions which are close, in an $L^2$-sense, to their slow counterparts.
All states and eigenfunctions are posed on a periodic domain of length $L=500$, so that for $\varepsilon > 0.01$ the stable wave is effectively isolated, while for $\varepsilon=0.002$ the stable pulse train does not relax to the quiescent state \CM{-- i.e. does not pass through the rest state -- while the unstable wave is isolated for all $\varepsilon$ considered}.

\begin{figure}[htpb]
	\includegraphics[width=\columnwidth]{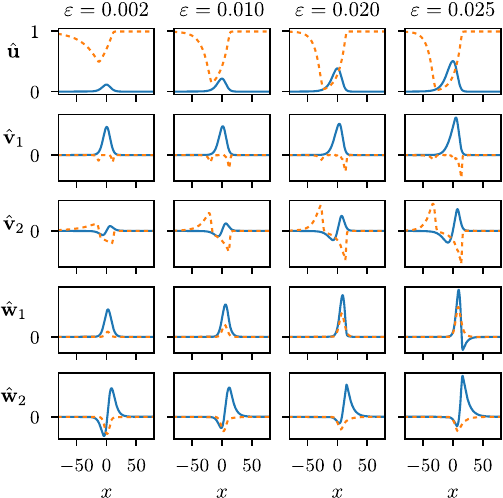}
	\caption{Linear theory ingredients for the Mitchell-Schaeffer model. (Top row) Unstable pulse solution $\hat{\vec{u}}$ for $\varepsilon \in \{0.002, 0.010, 0.020, 0.025\}$, with (second row)  $\hat{\vec{v}}_1$, (third row) $\hat{\vec{v}}_2$, (fourth row) $\hat{\vec{w}}_1$, and (fifth row) $\hat{\vec{w}}_2$. \CM{The first component of each variable is the solid (blue) curve, and the second component is shown as dashed (orange).}}
	\label{fig:9}
\end{figure}

Figure~\ref{fig:9} shows the linear theory ingredients for the Mitchell-Schaeffer model.
Surprisingly, the scaling of the leading eigenfunctions is still affected by the variation in $\varepsilon$, such that, for $\varepsilon=0.002$, similar considerations about the predictive power of the method should be taken into account relative to $\varepsilon=0.025$.
We supply the linear theory ingredients to the prediction program to estimate the necessary quenching amplitude for a variety of perturbation widths and positions\CM{, and likewise determine the critical quenching amplitude by direct numerical simulation for verification}.

\begin{figure*}
	\includegraphics[width=\textwidth]{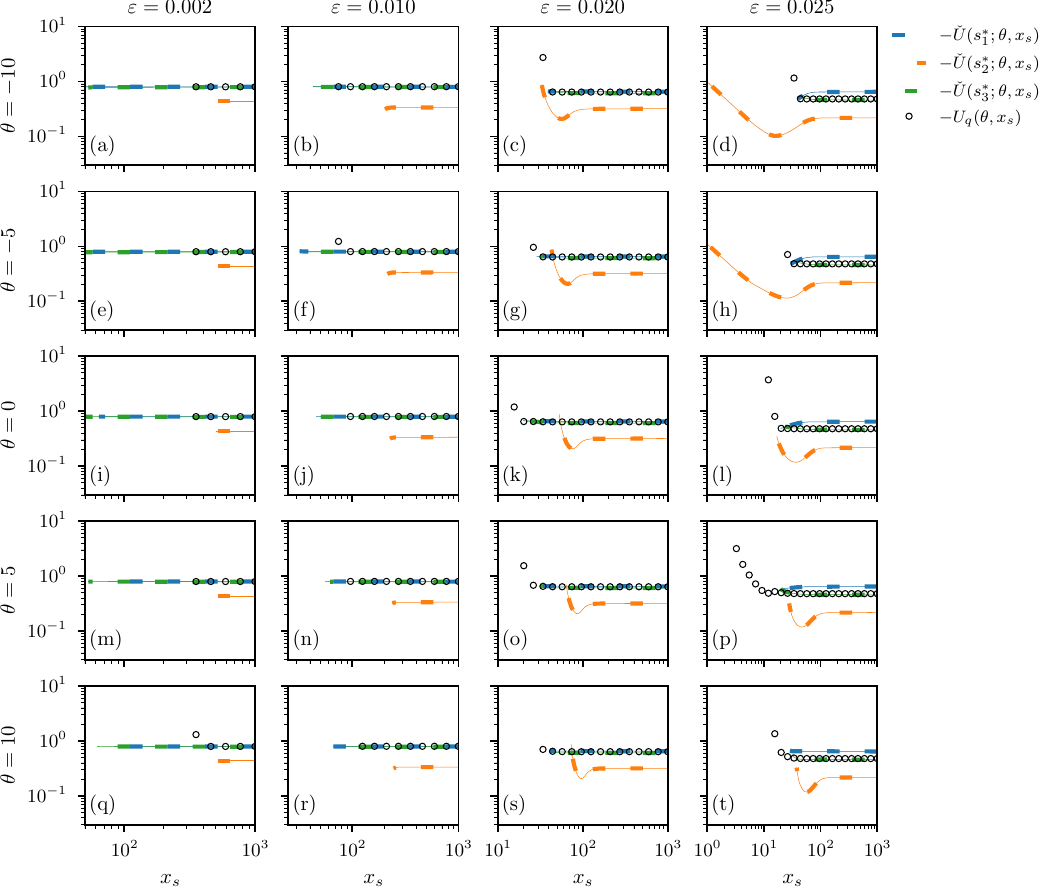}
	\caption{Critical quenching amplitude predictions \CM{for the Mitchell-Shaeffer model} using \CM{selectors} $\vmu{1}$ (blue), $\vmu{2}$ (orange), and $\vmu{3}$ (green) compared to the quenching amplitude computed using DNS of the perturbed stable pulse (black, $\circ$).}
	\label{fig:10}
\end{figure*}

The critical quenching predictions for the Mitchell-Schaeffer model are shown in figure~\ref{fig:10} for $\theta \in \{-10,-5,0,5,10\}$.
\CM{Similarly to figure~\ref{fig:7}, the} predictions transition from quasi-constant regions of perturbation amplitude when $x_s$ is large to quickly growing critical quenching amplitude for smaller widths.
The sensitivity to the position of the perturbation is more subtle in the Mitchell-Schaeffer model, i.e. we do not see substantial variation in the lower bound of $x_s$ across $\theta$, and the sensitivity to the difference between the fast and slow waves (i.e., $\varepsilon$) is more severe.
In particular, for small time-scale separation ($\varepsilon=0.025$) we observe a finite (soft) lower bound for $x_s$ in both the DNS and linear predictions, distinct from FitzHugh-Nagumo.
Likewise, the transition to growing quenching amplitudes for small $x_s$ is rarely relevant; instead nearly every prediction is within a tight tolerance of the asymptotic prediction, and the DNS quenching results only show growing quenching amplitudes for very small regions of the $(x_s,\theta)$-plane.

It is for the more similar pairs of waves (e.g., $\varepsilon=0.025$) that the difference between the heuristics is most clear, especially in the asymptotic $x_s\to\infty$ regime.
The predictions for large $x_s$ are only accurate for $\vmu{3}$; while $\vmu{2}$ systematically under-estimates the magnitude of the perturbation, and $\vmu{1}$ over-estimates the quenching amplitude specifically for large $x_s$.
Of particular note is that for $\varepsilon\geq0.020$, the transition from the asymptotic regime to larger quenching amplitudes is effectively captured by the linear theory (using $\vmu{3}$), such that the linear theory may accurately predict quenching applied solely to the wavefront, an essentially nonlinear effect.
The success of determining the transition length indicates that the linear prediction theory would benefit significantly from a consideration of higher-order effects, such that the applicability of the theory to larger time-scale separations could capture the linear-to-nonlinear transition for quenching, and predict successful quenching of the wavefront for ionic models.

\section{Discussion}

Some combinations of heuristic and parameters leads to considerably less accurate predictions; notably $\vmu{2}$ \CM{with $\gamma=0.01$ for the FitzHugh-Nagumo model (cf. figure~\ref{fig:7})}, where the prediction for the critical quenching amplitude differs by a factor of roughly $10\CM{\times}$ compared to \CM{$\vmu{1}$ and $\vmu{3}$}.
Consider the similarity of $\check{\vec{\Phi}}_3(x-s;s)$ and $\check{\vec{\Phi}}_2(x-s;s)$; i.e. the two correspond precisely for self-adjoint problems where $\hat{\vec{w}}_j \equiv \hat{\vec{v}}_j$.
With no further effort we may confidently blame the non-normality of the critical wave linearization and concomitant scaling of the components of the unstable pulse and eigenfunctions, \CM{similar to the ignition problem as $\gamma \to 0$~\cite{marcotte2020predicting}.}
We seek the dominant contribution to $\vmu{2}-\vmu{3}$ in the neighborhood of $s=s_3^*: \check{\mu}_3(s_3^*; \theta, x_s) = 0$.
As the perturbation envelope is unaffected by the details of this expansion, we consider the $\vPhi{l}$ directly,
\begin{align*}
    &\vPhi{2} - \vPhi{3} \\
    &= \check{\vec{\Phi}}_2(x-s_3^* - \delta; s_3^* + \delta) - \check{\vec{\Phi}}_3(x-s_3^* - \delta; s_3^* + \delta), \\
    &\approx \check{\vec{\Phi}}_2(x-s_3^*; s_3^*) \\
    &+ \delta \cdot \partial_s \left(\check{\vec{\Phi}}_2(x-s; s) - \check{\vec{\Phi}}_2(x-s; s)\right)|_{s=s_3^*} + O(\delta^2)\CM{,}
\end{align*}
\CM{where we have dropped the term $\check{\vec{\Phi}}_3(x-s_3^*; x_s)$ as it will not contribute to the inner product with the perturbation envelope, by construction.}
We elect to consider the $O(\delta)$ term directly, dropping the implicit dependence of the unstable pulse and eigenfunctions on $\xi=x-s$ \CM{where unambiguous}, and simplifying \CM{the derivative factor,}
\begin{align}\label{eq:DPhi}
    \begin{split}
    \delta^{-1}&\left(\vPhi{2} - \vPhi{3} - \check{\vec{\Phi}}_2(x-s_3^*; s_3^*)\right) \\
    &=\langle \hat{\vec{v}}_2 | \hat{\vec{u}} -\check{\vec{u}}(\xi+s_3^*) \rangle \hat{\vec{w}}_1\\
    &+ \langle \hat{\vec{w}}_2 - \hat{\vec{v}}_2 | \hat{\vec{u}} -\check{\vec{u}}(\xi+s_3^*)\rangle \partial_x\hat{\vec{w}}_1\\
    &+ \langle \hat{\vec{w}}_1 | \hat{\vec{u}} -\check{\vec{u}}(\xi+s_3^*) \rangle \partial_x (\hat{\vec{w}}_2 - \hat{\vec{v}}_2).
    \end{split}
\end{align}
Since we are interested in the perturbation magnitude in the \CM{large-width} limit, we take $\delta \equiv s_2^* - s_3^*$ and take the inner product \CM{of} $\vX{}$ with \eqref{eq:DPhi} \CM{in the large-width limit} $\lim_{x_s\to\infty}$,
\begin{align}\label{eq:Dmu}
    \begin{split}
    \lim_{x_s\to\infty}\delta^{-1}&\left( - \check{\mu}_3(s_2^*; \theta, x_s) - \check{\mu}_2(s_3^*; \theta, x_s) \right) \\
    =~&\langle \hat{\vec{v}}_2 | \hat{\vec{u}} -\check{\vec{u}}(\xi+s_3^*) \rangle \langle \hat{\vec{w}}_1 | \vX{} \rangle.
    \end{split}
\end{align}
That is, in the \CM{large-width} limit the inner products with differentials of the eigenfunctions become the eigenfunctions evaluated at their endpoints\CM{; for pulse solutions, these} are equal, leaving only the term proportional to $\hat{\vec{w}}_1$.
We estimate the magnitude of $\delta$ by inverting \eqref{eq:Dmu},
\[
    \delta \approx -\frac{\check{\mu}_3(s_2^*; \theta, x_s) + \check{\mu}_2(s_3^*; \theta, x_s)}{\langle \hat{\vec{v}}_2 | \hat{\vec{u}} -\check{\vec{u}}(\xi+s_3^*) \rangle \langle \hat{\vec{w}}_1 | \vX{} \rangle},
\]
and note that generically the numerator will be small (but typically non-zero), while the denominator will be large.
If we assume that $\partial_s\CM{\check{U}}(s_3^*) = 0$ (i.e., $\vmu{3}$ extremizes the perturbation amplitude), then
\begin{equation}\label{eq:deltadev}
    |\CM{\check{U}}(s_2^*) - \CM{\check{U}}(s_3^*)|\approx \frac{\delta^2}{2}|\partial_s^2\CM{\check{U}}(s_3^*)|,
\end{equation}
which may be \CM{comparable to $|\CM{\check{U}}(s_3^*)|$}, even if $\delta$ is small (cf. Appendix~\ref{sec:app} for numerical details).
\CM{We may then conclude that for a severely non-normal linearization about the unstable pulse -- the generic case for excitable media models -- heuristics which do not rely on the right eigenspace are more reliable indicators for frame selection in the quenching predictions.}

\CM{The potential for the techniques developed here to improve the efficacy of multi-phase defibrillation approaches should be considered carefully.
In principle, a technique which affects the \emph{depolarized} tissue directly during early defibrillation phases -- when depolarized tissue makes up the majority of the tissue -- should improve the efficacy of initial stimulus pulses aimed at restoring the tissue to a quiescent state, and thus re-establishing the normal rhythm.
However, the results presented in this study indicate a particular weaknesses of this approach: that the energy cost to quenching an excitation wave is substantially higher than the energy cost of ignition.
Therefore, quenching requires both a larger energy cost per excitation wave, and over a larger tissue region, and thus may require significantly more energy for effective use than current ignition-based multi-phase defibrillation techniques, i.e. Low-Energy Atrial Pacing (LEAP)~\cite{fenton2009termination,luther2011low}.
A thorough accounting of the energy requirements for the present method in a realistic defibrillation setting is beyond the scope of this study, but is nonetheless essential for determining the relevance of the method in a clinical setting.}

A similar quenching effect was investigated numerically in the modified three-component Oregonator, and experimentally using optical control methods in the Belousov-Zhabotinskii reaction\CM{~\cite{nishi2017achilles}}.
That work considered specifically time-distributed controls in the co-moving frame of the pulse, i.e. those of form $\check{\vec{X}}^\prime(x-\theta-\check{c}t; x_s)\Theta(t)$, which differs from those investigated in this \CM{study} $\vX{}\delta(t)$.
Nonetheless, \CM{the ingredients are the same}; their `dev' is analogous to our $\theta$, their $d_\mathrm{irrad}$ is analogous to our $x_s$, and their $\varepsilon_\mathrm{irrad}$ is analogous to our $|U_q|$.
\CM{A direct comparison requires an extension of the quenching theory to co-moving perturbations.}

Finally, we return to the question of how to best measure the inter-state distance for the selection of an appropriate frame; i.e. how should we construct $\vmu{l}$ using the ingredients of the stable and unstable state for the purposes of predicting quenching?
In contrast to the ignition case, for quenching we can make a strong argument for the third heuristic, which chooses a frame in which the projection onto the center-stable eigenmode of the unstable pulse is identically zero.
It is simple to see that, if the $L^2$-distance is already small---when the fast and slow pulses are similarly shaped---then $\check{\mu}_3$ determines a frame which is close to the one in which the perturbation \CM{minimally excites} the translational mode of the \emph{stable} wave in addition to the unstable wave \CM{thus} the perturbation generated from the third heuristic may be seen as more efficient for quenching.
However, the numerical evidence suggests that no single frame selector is uniquely prescient---for FitzHugh-Nagumo and $\gamma > 0.01$, all three predictors give nearly indistinguishable results for all cases; while for $\gamma=0.01$ \CM{and the Mitchell-Schaeffer model}, $\vmu{2}$ is uniquely poor.
For $\gamma=0.001$, no heuristic accurately captures the lower bound of $x_s$ for effective quenching observed in the \CM{DNS}.
This suggests a different criterion by which to consider the frame selectors: \CM{\emph{what heuristic produces the fewest smaller-amplitude predictions in the parameter ranges over which DNS results show that quenching is possible}}?
For \CM{the FitzHugh-Nagumo and Mitchell-Schaeffer models}, $\vmu{2}$ is unambiguously worst by this metric, as it systematically underestimates the quenching amplitude.
To put such predictions into practice would fail to \CM{quench} an excitation---which in a medical context could be disastrous.

\section{Conclusions}

Given a stable propagating solution and a family of perturbations parameterized by their width $x_s$ and position $\theta$, we are able to predict the critical quenching amplitude which corresponds to the cessation of propagation.
The method is effective in the archetypal FitzHugh-Nagumo model with isolated pulse solutions and the Mitchell-Schaeffer model with pulse train solutions.
We find that the linear theory achieves qualitative accuracy with minimal insight, and quantitative accuracy in regimes where the linear assumptions of the perturbation theory are valid and the leading adjoint eigenfunctions are `small' in a sense preserved by the norm.
The method is sufficiently fast and parsimonious to be useful in optimization searches, and thus opens new avenues for the selection of \emph{optimal} positions $\theta$ and widths $x_s$ in addition to the prediction of the critical amplitude for such pairings.
Finally, we find the predominant shortcoming of the application of the linear theory is the determination of the \emph{correct} reference frame, as no generally applicable uniqueness results are available, though in practice we resolve such difficulties through continuation.
We expect that additional heuristics, in analogy with the critical ignition problem, may be reasonably considered, but have focused primarily on the numerical results in this \CM{study}.

We have not treated the construction of optimal perturbations in this \CM{study}, in either the linear~\cite{farrell1988optimal} nor the nonlinear~\cite{pringle2010using} cases.
Likewise, we have not considered the more general case of optimal control of cardiac excitation patterns using, e.g., adjoint flow optimization of a stimulating current or applied electric field~\cite{chamakuri2015application}.
\CM{However, we may make some general comments on the relative cost of the approach detailed in this study compared to these alternative methods.}
\CM{The construction of the linear  optimal perturbations} requires a significant number of \CM{leading} eigenmodes and subsequent formation of all inner products of these modes for the prediction of transient amplification~\CM{\cite{whidborne2011computing}}, while the nonlinear optimal perturbation requires \CM{the solution of the adjoint flow in concert with the forward nonlinear PDE in a repeated optimization pass}~\cite{lecoanet2018connection}.
\CM{In contrast, for} this study only \CM{two pairs of} leading eigenmodes are \CM{needed for the calculation of the quenching perturbation, and we need only solve the underlying PDE system in cases of verification -- i.e., sparingly}.
\CM{The only remaining computational cost of the method is the calculation of the fast Fourier Transform, i.e. comparable to a single time-step in the PDE solve.}
\CM{In the most generous case, we estimate that the method proposed in this study represents a computational savings over the linear and nonlinear optimal perturbation approaches of skipping all eigenfunction calculations after the first two, or skipping the entire PDE solve and optimization loop after the first time-step, respectively.}
\CM{The parsimony of the method yields a significant reduction in computational cost, requiring fewer eigenmodes or fewer PDE solves, making it an asset for quenching predictions.
The efficiency of the method presents opportunities for embedding these calculations into larger or on-line computations, e.g. the determination of model parameter values driven by observations of real tissue.}

\CM{We consider two natural extensions of the present work.}
\CM{First, an extension of the current approach} to account for front solutions---solutions $\check{\vec{u}}$ of equation~\eqref{eq:bvp} where $\lim_{x\to+\infty}\check{\vec{u}}(x) \neq \lim_{x\to-\infty}\check{\vec{u}}(x)$.
The asymptotic arguments \CM{presented in this study} rely on the far-field \CM{values of the eigenfunctions and state; since front solutions also have reasonably localized leading eigenfunctions, it is only the latter presents an issue for the theory.}
\CM{Indeed much of the numerical work may be trivially extended using fast algorithms for the computation of the Fourier extension~\cite{matthysen2016fast} and enable the treatment of front quenching similarly to pulse quenching.}

\CM{Second, the application of the present theory to excitable media with heterogeneity -- i.e. most physical systems.}
Without a homogeneous medium, \CM{the system does not possess translational symmetry and the selection of the reference frame loses meaning.}
\CM{However, cardiac systems exhibit very little} sensitivity to perturbations far from the salient features of the state; in the one-dimensional context this is the critical wave peak, and in higher dimensions this is a topological feature of the state (i.e. spiral origins or scroll wave filaments)~\CM{\cite{biktasheva2003wave, biktasheva2006localization, marcotte2015unstable}}.
Similar considerations for the sensitivity of spiral waves to localized medium heterogeneity~\CM{\cite{biktasheva2009computation, biktashev2011evolution, kharche2015computer}} suggest this hurdle can be \CM{overcome by treating `small' deviations from homogeneity as perturbative.}

\CM{The results of this study indicate that the quenching of pulses in excitable media -- which are not amenable to investigation through direct linearization -- are nonetheless predictable with a small amount of model insight.
Additionally, we have shown that the predictions may be evaluated efficiently (especially compared to competing general techniques for the prediction of transitions in nonlinear systems) and accurately, to determine the necessary conditions of quenching excitable pulses.
Finally, we have argued that quenching may be relevant for realistic multi-phase defibrillation techniques, though the results indicate that the energy costs of the approach may be prohibitive compared to existing ignition-based approaches.}

\section{Acknowledgements}

CDM would like to thank Prof. Vadim Biktashev (University of Exeter) for helpful discussions.

\appendix
\section*{Appendix}

\subsection{Traveling wave continuation \& eigenproblems}

As mentioned in the main text, we compute families of excitation pulses using \textsc{Auto-07p}~\cite{doedel2007auto}.
This involves casting \eqref{eq:ivp} into first-order form in $u_1$, $u_2$, and $u_3\equiv\partial_xu_1$:
\begin{align}
	\begin{split}\label{eq:fos}
	u_1' &= u_3,\\
	u_2' &= -f_2(u_1,u_2)/c,\\
	u_3' &= -(c u_3 + f_1(u_1,u_2) + J)/D,
	\end{split}
\end{align}
where $c>0$ and $D>0$.
An initial condition corresponding to the quiescent state, $\bar{\vec{u}}$, appended with an extra zero corresponding to the first derivative of $\bar{u}_1$, serves as initial condition.
The system is then continued in the stimulation current $J$ to a Hopf bifurcation, at which point the solver switches to the family of periodic orbits emanating from the bifurcation, now in $(J,c)$.
When states of sufficient length are found, the periodic solutions are continued until $J=0$, and the resulting solution serves as a starting point for the continuation to generate an interesting family of solutions; e.g., for FitzHugh-Nagumo, in $(\gamma,c)$ as depicted in \CM{figure~\ref{fig:3}}.
In the experiments presented in this work, we use $\mathtt{NTST}=1000$ and $\mathtt{NCOL}=4$, leading to a $4001$-element discretization of the pulse at each $(\gamma,c)$-pair.
The solutions generated by this process are then written to file, along with their parameters.

We use the \textsc{Auto-07p} generated solutions as initial estimates for a boundary value problem (BVP) posed on a Chebyshev domain with $M$ basis modes, and $N=2M$ grid points; when $N > M$, the calculation of nonlinear terms is more accurate due to dealiasing of the underlying nodal integrals.
The boundary value problem (BVP) is constructed in \textsc{Dedalus2}~\cite{burns2020dedalus}, \CM{as well as} the eigenvalue (EVP) and initial-value problems (IVP).
As \textsc{Dedalus2} requires casting the system to first order in Chebyshev-basis spatial derivatives, we arrive at system \eqref{eq:fos}; as a third order system in two (primary) variables, in order to avoid asymmetric boundary conditions (which will, typically, pollute the eigenspectrum of the pulse)~\cite{driscoll2016rectangular,aurentz2017block}.
We shall also note that the implementation in \textsc{Dedalus2} uses an implicit \CM{tau} method, which is equivalent to dropping the rows in the discretization corresponding to the highest frequency modes in the Chebyshev expansion and replacing those rows with the boundary conditions.
Thus our BVP, EVP, and IVP solutions are technically only correct to the leading $M-3$ modes.
More general tau methods are available in \textsc{Dedalus3} through explicit inclusion of tau terms~\cite{dedalus3tau}.

The boundary conditions used in the solution of this BVP are projective~\cite{marcotte2020predicting}, which permits the finite expansion to approximate the homoclinic orbit rather than the periodic wave train solution (which would be used with the periodic boundary conditions of the \textsc{Auto-07p} periodic orbits).
We refine the solution on the dealiased Chebyshev grid using a Newton solver for the system until the maximum of the Newton updates is smaller than $5 \times 10^{-15}$ and the Newton update contains all the energy in the dealiased modes (i.e., modes larger than $M$).
For the FitzHugh-Nagumo model, $M = 2^{12}$, which we found to be sufficient on a domain of size $L=2700$.
Likewise, as the domain is significantly larger than the critical pulse, the projective boundary conditions effectively satisfy periodic boundary conditions up to an error in the deviation of the endpoints from the rest state; in our experience, this is smaller than $10^{-13}$, but there is no known guarantee for the exactness of this approximation for the Chebyshev-T expansion on Gauss-Quadrature nodes.

The refined BVP solution is then passed to an EVP solver, which constructs the linearized system about the solution using (analytical) expressions for $f_{ijk} \equiv \partial_i\partial_jf_k$, $i,j,k \in \{1,2\}$:
\begin{align}
	\begin{split}
	0 &= \sigma v_1 - (D v_3' + c v_3 + f_{101} v_1 + f_{011} v_2), \\
	0 &= \sigma v_2 - (       + c v_3 + f_{102} v_1 + f_{012} v_2), \\
	0 &= v_3 - v_1',
	\end{split}
\end{align}
for the eigenvalue $\sigma$, on the same Chebyshev domain as was used in the BVP (with the datatype elevated to \texttt{complex128}, to account for complex-valued modes).
The EVP solver likewise inherits the projective boundary conditions from the BVP solver.
However, the EVP solver may be used to compute the left or right eigenmodes by appropriate substitutions in the formation of the eigenproblem, especially the boundary conditions.

Finally, we consider the IVP used in the direct numerical simulation of perturbed stable waves.
Since we already have a preponderance of solutions computed using a finite-difference method and adaptive time-stepping, we treat the spectral solution of the IVP as a verification of the results.
This calculation reuses the same the $M$-mode solutions of the BVP solver to form the initial condition.
In general, one should treat the extension from the BVP solution with projective boundary conditions to the IVP with periodic boundary conditions carefully.
The former is a segment of the infinitely long homoclinic orbit solution, while the latter is a periodic orbit ``close'' to the homoclinic for sufficiently large $L$, $M$, and $N$.
In practice, we treat the extension as both a check that we have sufficiently resolved the tails of the solution in the homoclinic case (by comparing the difference in the state variables at the endpoints), and an effective limit to our solution accuracy for the periodic problem.
The initial condition is then perturbed according to the perturbation bisection problem detailed in the text, which makes the initial condition non-smooth; for this reason we do not dealias the state for the IVP ($N=M$) to account for the sharp features of the perturbations to the state for the quenching problem.
Given the non-smooth initial condition, we found it was necessary to make small timesteps initially, which we manage by multiplying the base time step \CM{$\delta t$} by a power of $2$ based on the progress of the simulation.
That is, at time $0 \leq t_i \leq T$, $\CM{\log_2\left((t_{i+1}-t_i)/\delta t\right)} = \min([\max([\ceil{\log_2(t_i/T + 2^{-12})},-12]),0])$; so initially, the simulation makes $2^{12}$ times more time-steps per unit time than at the end of the simulation.
This stabilizes the already strongly dissipative SBDF4 method~\cite{ascher1997implicit} when \CM{$U_q$ is large and negative, i.e., $U_q\sim-10^{2})$}.
The root-finding problem for the amplitude is solved to a tolerance of $10^{-10}$ for each $(x_s,\theta)$ pair by repeatedly calling the IVP with perturbed initial conditions within a bisection procedure, with early exit \CM{from the IVP} in the case that $\|\vec{u}(t,x)- \bar{\vec{u}}\|_1 < 10^{-10}$ and $t < T = L/(2\check{c})$.

\subsection{Root-finding and continuation\label{sec:app}}

As the determination of the optimal frame is posed in terms of a scalar root-finding problem, cf. \eqref{eq:mu}, and $\check{\mu}_l(s; \theta, x_s)$ is generally an unknown but very complicated function, significant care in the identification, refinement, and interpretation of roots must be taken.
Of particular relevance for the predictions of the linear theory, we have no strong guarantees of uniqueness for the roots of $\check{\mu}_l(s; \theta, x_s)$ for every given perturbation envelop $\check{\vec{X}}(x-\theta; x_s)$, nor for any model-supplied $\hat{\vec{u}}(\xi)$, $\hat{\vec{w}}_1(\xi)$, $\hat{\vec{w}}_2(\xi)$, and $\check{\vec{u}}(\xi+s)$.

As part of the published linear theory code~\cite{gitrepo}, we construct $\check{\mu}_l(s; \theta, x_s)$ two ways: using the convolution approach detailed in the main text, and through explicit reconstruction and shifting of \CM{$\check{\vec{\Phi}}_l(x-s, s)$} to form
\begin{equation}\label{eq:tol}
	\check{\mu}_l^{!}(s; \theta, x_s) \equiv \langle \CM{\check{\vec{\Phi}}_l(x-s, s)} | \check{\vec{X}}(x-\theta; x_s) \rangle
\end{equation}
for \CM{$2^8$} test shifts \CM{sampled uniformly from $[0, L)$}.
Deviations are on the order of $|\check{\mu}_l(s; \theta, x_s) -\check{\mu}_l^{!}(s; \theta, x_s)| \lesssim 10^{-6}$ for the problems considered, which is sufficient for the linear theory predictions; we note the maximum error as $\mathtt{tol}$ such that $1 \gg \mathtt{tol} > 0$.

Indeed, the leading issue with the direct application of the linear theory (instead of the continuation from \CM{asymptotically large widths}) is not the inaccuracy of scalar root-finding (this would imply that the optimal frame is adjacent to extremely sub-optimal frames), but rather the often extremely large number of roots computed resulting from oscillations in the input fields.
The linear prediction code interpolates the input fields onto a Fourier grid of fixed size to speed up the convolutions; typically $2^{13}$ is sufficient for our purposes, as the number of grid points only affects the resolution of the predictions for very small perturbation widths.
However, the Fourier expansion guarantees that the number of roots of $\check{\mu}_l(s; \theta, x_s) = 0$ is an even number.
For the $2^{13}$ Fourier expansion, we will typically find between $2$ and $2\times 10^3$ prospective roots without additional insight; thus the \CM{near-}uniqueness of the asymptotic perturbation is essential for tractability.

Unfortunately, the continuation requires the solution of all intermediate problems for small perturbation widths, which is inefficient compared to a direct computation for any single $x_s$.
In principle, the direct computation may compute the correct asymptotic branch, but does not determine it uniquely; it is unclear whether a filtering process applied to the roots specifically may discriminate the asymptotic branch without computing the asymptotic solution.
For numerical reasons, the direct (non-continuation) calculation requires the introduction of filters based on the value of $\mathtt{tol}$ determined by the comparison of \CM{$\check{\mu}_l^{!}(s; \theta, x_s)$} and $\check{\mu}_l(s; \theta, x_s)$, c.f. discussion surrounding \eqref{eq:tol}.
\CM{Often} distinct roots yield very similar traces (recall that for each $(x_s,\theta)$, $\Us$ is a one-dimensional function of $s$, cf. figure~\ref{fig:7} \CM{\& \ref{fig:11}}).
Finally, we note that none of the filtering steps explicitly filter on the sign of the perturbation -- each frame selector will, in most circumstances, have roots which correspond to \emph{positive}-valued quenching perturbations.
\CM{We refer the interested reader to the public repository and comments therein.~\cite{gitrepo}}

We initialize the continuation problem seeking solutions $(x_s, \Us)$ while controlling $x_s$, where $\theta$ is held constant.
The continuation problem begins from the asymptotic case so that we may identify \CM{$\lim_{x_s\to\infty}\Us$} uniquely.
By treating the continuation with natural parameterization (i.e., not pseudo-arclength) we do not track solutions around curves where $x_s' = 0$; this is desirable since reversals in the branch would make the prediction of the critical amplitude non-unique for some $x_s$.

\begin{figure*}[htb]
    \centering
    \includegraphics[width=\textwidth]{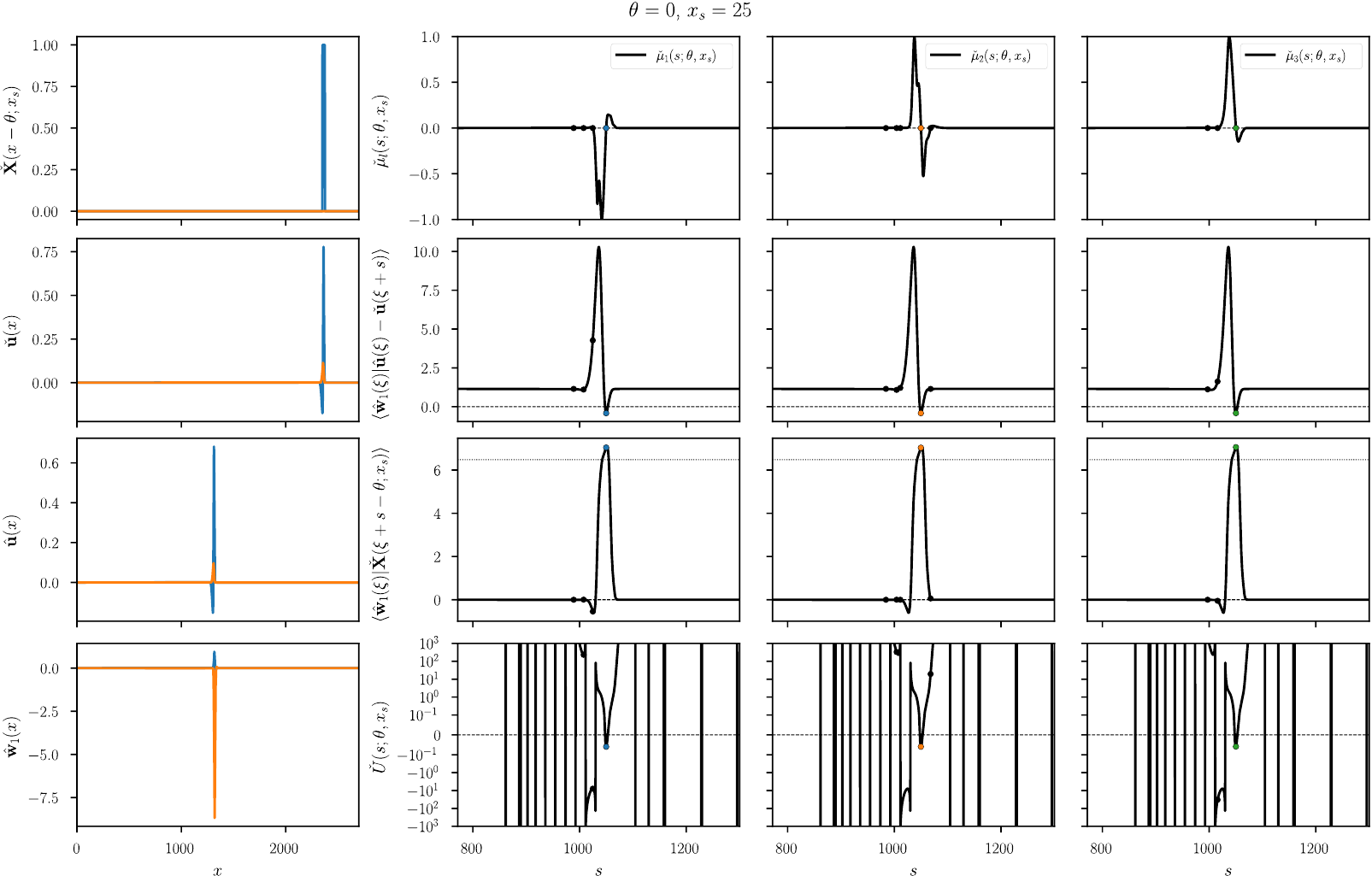}
    \caption{Continuation to $x_s = 25$ for the quenching problem with $\theta=0$ applied to the FitzHugh-Nagumo pulse with $\gamma=0.025$. Note in the depiction of the denominator of $\Us$ that the traced root is larger than the asymptotic (dotted line).}
    \label{fig:11}
\end{figure*}

An additional benefit of the continuation \CM{approach} is the comparison of sub-expressions from the asymptotic case to smaller perturbation widths.
In figure~\ref{fig:11}, we show a diagnostic figure generated during the continuation process, which shows the perturbation, stable and unstable solutions, and the leading unstable left eigenfunction (left-most column), and a four-by-three block of diagnostic computations from the continuation.
In the remainder of the figure we show $\vmu{l}$ for each $l$ (line, top row), and the roots of this function (black dots), while highlighting the asymptotic branch (colored dot).
The second row shows the numerator of $\Us$, likewise highlighting the roots of $\vmu{l}$.
The third row shows the denominator of $\Us$, highlighting the roots of $\vmu{l}$ and the asymptotic value.
The final row shows the value of $\Us$, highlighting the roots, and oscillating wildly for $x_s \ll L$ due to the division by near-zero values, i.e. $\langle \hat{\vec{w}}_1(\xi) | \check{\vec{X}}(\xi+s-\theta; x_s) \rangle \approx 0$.
An animation of the continuation sequence is available {\href{https://durhamuniversity-my.sharepoint.com/:v:/g/personal/tkdn52_durham_ac_uk/ESV53d0r_GxHtyuN2ucP7O0BEvMnajGry86kQi_Dbj0K0A?nav=eyJyZWZlcnJhbEluZm8iOnsicmVmZXJyYWxBcHAiOiJPbmVEcml2ZUZvckJ1c2luZXNzIiwicmVmZXJyYWxBcHBQbGF0Zm9ybSI6IldlYiIsInJlZmVycmFsTW9kZSI6InZpZXciLCJyZWZlcnJhbFZpZXciOiJNeUZpbGVzTGlua0NvcHkifX0&e=5qB4cH}{here}}.
However, in practice the asymptotic root is far from the regions where the denominator of \eqref{eq:U} oscillates around $0$, by construction, and the predictions do not suffer from numerical inaccuracies because of these far-field effects.
Rather, as explained in the main text \eqref{eq:deltadev}, we find some sensitivity to the curvature of $\Us$ for $\vmu{2}$ about $s_3^*$; estimates of this factor are $|\partial_s^2 U(s_3^*)| \sim 10^{+2}$ and $-10^{+5}$ in some cases.

\section*{References}
\bibliography{quenching}

\end{document}